\documentclass[10pt,conference]{IEEEtran}
\IEEEoverridecommandlockouts
\usepackage{graphicx}
\usepackage{subfigure}
\usepackage{cite}
\usepackage{bm}
\usepackage{amsmath}   % From the American Mathematical Society
\usepackage{amsthm}
\usepackage{cases}
\usepackage{amsfonts,amssymb}
\usepackage{cuted}\stripsep-3pt
\usepackage{lipsum}
\usepackage{algorithm}
\usepackage{algorithmic}
\usepackage{stfloats}
\usepackage[absolute,showboxes]{textpos}

\TPMargin{3pt}

\newcommand{\copyrightstatement}{
    \begin{textblock}{0.84}(0.08,0.95) % tweak here: {box width}(left position, bottom position)
         \noindent
         \footnotesize
         \copyright 2022 IEEE. Personal use of this material is permitted. Permission from IEEE must be obtained for all other uses, in any current or future media, including reprinting/republishing this material for advertising or promotional purposes, creating new collective works, for resale or redistribution to servers or lists, or reuse of any copyrighted component of this work in other works. DOI: 10.1109/GLOBECOM48099.2022.10001218.
    \end{textblock}
}
\ifCLASSINFOpdf
\else
\fi
\begin{document}
\copyrightstatement
\title{Joint Reflection and Power Splitting Optimization for RIS-assisted OAM-SWIPT}


%\author{\IEEEauthorblockN{Runyu Lyu, \textit{Student Member, IEEE}, Wenchi Cheng, \textit{Senior Member, IEEE}, and Wei Zhang, \textit{Fellow, IEEE}~}% <-this % stops an unwanted space

\author{\IEEEauthorblockN{Runyu Lyu and Wenchi Cheng}% <-this % stops an unwanted space
\IEEEauthorblockA{State Key Laboratory of Integrated Services Networks, Xidian University, Xi'an, China}
\vspace{-20pt}
\thanks{This work was supported in part by the National Key Research and Development Program of China under Grant 2021YFC3002102 and in part by the Key R\&D Plan of Shaanxi Province under Grant 2022ZDLGY05-09.
%This work was supported by the National Natural Science Foundation of China (No. 61771368), Foundation of CETC Key Laboratory of Data Link Technology (CLDL-20182411), Key Area Research and Development Program of Guangdong Province under grant No. 2020B0101110003, and in part by Shenzhen Science \& Innovation Fund under Grant JCYJ20180507182451820.
%
%Runyu Lyu and Wenchi Cheng are with the State Key Laboratory of Integrated Services Networks, Xidian University, Xi’an, 710071, China (e-mails: rylv@stu.xidian.edu.cn; wccheng@xidian.edu.cn).
%%
%%W. Zhang is with the School of Electrical Engineering and Telecommunications, the University of New South Wales, Sydney, NSW, Australia (email: w.zhang@unsw.edu.au).
}
}

%{Shell \MakeLowercase{\textit{et al.}}: Bare Demo of IEEEtran.cls for IEEE Transactions on Magnetics Journals}



% for Transactions on Magnetics papers, we must declare the abstract and
% index terms PRIOR to the title within the \IEEEtitleabstractindextext
% IEEEtran command as these need to go into the title area created by
\IEEEtitleabstractindextext{%
%\vspace{-10pt}
\begin{abstract}
Simultaneous wireless information and power transfer (SWIPT) can enhance the spectrum and power efficiencies of wireless communications networks. Line-of-sight (LOS) transmission is a typical SWIPT scenario. However, the strong channel correlation limits the spectrum and energy efficiencies of SWIPT in the LOS channel. Due to the orthogonal wavefronts, orbital angular momentum (OAM) waves can facilitate the SWIPT in LOS channels. With the assistance of the reconfigurable intelligent surface (RIS), both the energy efficiency and capacity can be further improved for the OAM-SWIPT systems. In this paper, we model the RIS-assisted OAM-SWIPT transmission and derive the optimal reflection coefficients and power splitting ratio for it. We first give the system and channel models. Then, we propose the transmission scheme. Based on the transmission scheme, we formulate the capacity and energy harvesting (EH) trade-off problem. We solve the problem by developing an alternating optimization algorithm. Simulations validate the capacity and EH enhancements brought by the RIS for OAM-SWIPT.
\end{abstract}

%\vspace{-5pt}
}

\maketitle

\IEEEdisplaynontitleabstractindextext

% \fi
\IEEEpeerreviewmaketitle

\vspace{-5pt}
%\addtolength{\textheight}{-0.44 cm}
\section{Introduction}
\IEEEPARstart{W}{ith} the emergences of Internet of Everything (IoE) system as well as applications such as virtual reality, brain-computer interfaces, and autonomous drone swarms\cite{6G}, the demand for wireless capacity will continue to grow. Also, more base stations, more complex networks, and the boosting of wireless devices increase the power consumption of the fifth-generation (5G) wireless communication as well as next-generation mobile networks\cite{6G_OAMin}. Therefore, green technologies that can improve both spectrum efficiency and energy efficiency are urgently needed. The radio-frequency (RF) signal can be harvested to achieve simultaneous wireless information and power transfer (SWIPT)\cite{SWIPT}, which can improve the performance of information networks in terms of spectrum efficiency and energy efficiency\cite{SWIPT_mag}.

\begin{figure*}[htbp]
\centering
%\vspace{-10pt}
\includegraphics[scale=0.52]{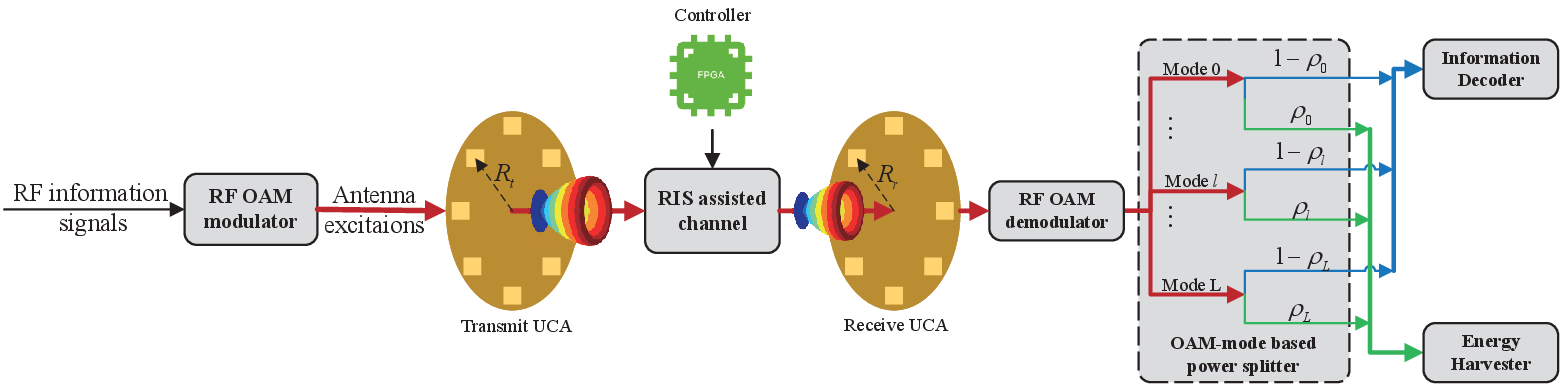}
%\vspace{-15pt}
\caption{The RIS-assisted OAM-SWIPT system model.} \label{fig:system_model}
\vspace{-15pt}
\end{figure*}
% 介绍涡旋和其在SWIPT的应用
Considering the energy reception efficiency, line-of-sight (LOS) RF transmission is a typical SWIPT scenario. However, the strong channel correlation of LOS channels limits the capacity and the application of SWIPT technology. Fortunately, orbital angular momentum (OAM)\cite{oam_light} waves can be multiplexed/demultiplexed together to increase the spectrum efficiency without additional power in LOS channels\cite{oam_low_freq_radio,OAM_NFC_mine}. Thus, OAM technology can increase energy efficiency and facilitate SWIPT in LOS channels\cite{SWIPT_OAM_mine}. However, there is no related works considering optimizing parameters such as the power splitting ratio of information decoding (ID) and energy harvesting (EH) for OAM based SWIPT\cite{SWIPT_OAM2}.

%However, few studies considered using OAM for SWIPT\cite{SWIPT_OAM2}, not to mention optimizing the power splitting ratio of information decoding (ID) and energy harvesting (EH).

% 扯上RIS
Reconfigurable intelligent surface (RIS) is also an appealing technique for next-generation communication systems because of its high cost-effectivity and energy efficiency\cite{Intelligent_Surfaces}. Combined with SWIPT technology, RISs can increase the EH efficiency\cite{IRS_MIMO}. Combined with OAM technology, RIS can be used to reflect the OAM waves blocked by obstacles and construct a direct path for the OAM-based SWIPT\cite{oam_irs}. Therefore, the RIS-assisted OAM-SWIPT is a promising technology for future energy-efficient and spectrum-efficient wireless communication.

In this paper, we model the RIS-assisted OAM-SWIPT transmission and derive the optimal reflection coefficients and power splitting ratio for it. We first build up the system and transmission models. Then, we formulated the capacity maximization problem subject to the EH requirement. We solved the problem by dividing it into two subproblems and alternatively solve them until convergence. Next, we validate the convergence of our developed algorithm via simulations. Simulations also validate the capacity and EH enhancements brought by the RIS for OAM-SWIPT. Additionally, the impact of the distance between the transmit UCA and the RIS are also simulated.

The rest of the paper is organized as follows. Section~\ref{sec:System_model} gives the system and channel models. Section~\ref{sec:Transmission_scheme} proposes the RIS-assisted OAM transmission scheme. In Section~\ref{sec:Problem_formulation}, we formulate the capacity maximization problem. Section~\ref{sec:Algorithm} solves the problem by giving an alternating optimization algorithm. In Section.~\ref{sec:Simulation}, simulations are given to validate the convergence of the algorithm as well as the capacity and EH enhancements brought by the RIS for OAM-SWIPT. The conclusion is given in Section~\ref{sec:Conclusion}.

\begin{figure*}[!bp]
\hrulefill% 一条分割线，长度为页面宽度
\setcounter{equation}{0}
\begin{subequations}
\begin{numcases}{}
\boldsymbol{\mathrm p}^{t}_{n_t} = \left[R_t\cos\left(\frac{2\pi \left(n_t-1\right)}{N_t}\right),R_t\sin\left(\frac{2\pi \left(n_t-1\right)}{N_t}\right),0\right];\\ \nonumber\\
\boldsymbol{\mathrm p}^{r}_{n_r} = \left[d_x,d_y,D\right] - R_r\cos\left(\frac{2\pi \left(n_r-1\right)}{N_r}\right)\frac{\boldsymbol{\mathrm b}^r}{\Vert\boldsymbol{\mathrm b}^r\Vert} + R_t\sin\left(\frac{2\pi \left(n_r-1\right)}{N_r}\right)\frac{\boldsymbol{\mathrm a}^r}{\Vert\boldsymbol{\mathrm a}^r\Vert};\\ \nonumber\\
\boldsymbol{\mathrm p}^{I}_{n_I} = \left[p_x,p_y,p_z\right] - \left[\left(n_I\ {\rm mod}\ N^c_I-1\right)d-\frac{d(N_I^c-1)}{2}\right]\frac{\boldsymbol{\mathrm b}^I}{\Vert\boldsymbol{\mathrm b}^I\Vert} + \left[\left(n_I\ |\ N^c_I\right)d-\frac{d(N_I^r-1)}{2}\right]\frac{\boldsymbol{\mathrm a}^I}{\Vert\boldsymbol{\mathrm a}^I\Vert}.
\end{numcases}\label{eq:coordinate}
\end{subequations}
%\hrulefill% 一条分割线，长度为页面宽度
\end{figure*}
\section{System and Channel Models}\label{sec:System_model}
\subsection{System Model}
Figure~\ref{fig:system_model} shows the RIS-assisted OAM-SWIPT system model, where the transmit and receive antennas are uniform circular arrays (UCAs) with $N_t$ and $N_r$ elements, respectively. The radii of the transmit and receive UCAs are denoted by $R_t$ and $R_r$, respectively. In this paper, we use UCAs as the transceiver antennas because UCA can generate waves with multiple OAM-modes digitally and simultaneously. However, other kinds of antennas, such as the spiral phase plate and metasurface, can also be used to generate OAM beams. The multi-stream RF information signals are first modulated into the excitations corresponding to the $N_t$ antenna elements. Then, the excitations of multiple OAM-modes are emitted via the transmit UCA and form the OAM waves\cite{oam_low_freq_radio}. After passing through the RIS-assisted OAM-SWIPT channel, the RF receive signals are first demodulated into the signal corresponding to each OAM-mode. Then, each OAM-mode signal splits into the ID stream and the EH stream using power dividers with arbitrary power division\cite{ArbitrarDivisionPowerDivider}. Combined with the RIS, the maximum data rate can be achieved under the minimum EH constraint by optimizing the reflection coefficients and the power splitting ratio of ID and EH.

\subsection{Channel Model}
\begin{figure}[htbp]
\centering
%\vspace{-10pt}
\subfigure[Channel model.]{
\begin{minipage}{1\linewidth}
\centering
\includegraphics[scale=0.63]{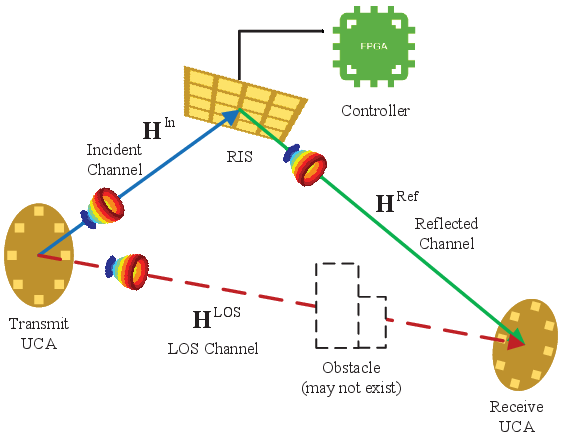}
\label{fig:channel_model}
\vspace{-10pt}
\end{minipage}
}\\
\subfigure[Coordinate system.]{
\begin{minipage}{1\linewidth}
\centering
\includegraphics[scale=0.63]{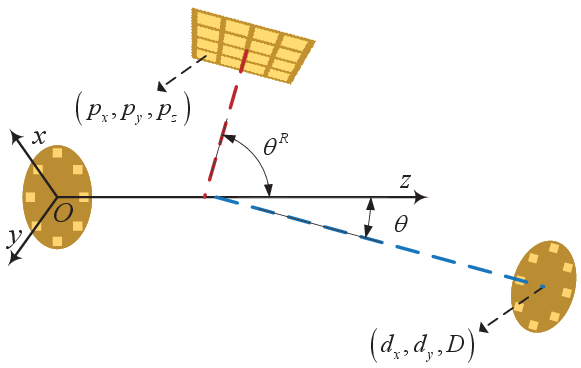}
\label{fig:channel_model2}
%\vspace{-3pt}
\end{minipage}
}
\centering
\caption{RIS-assisted OAM-SWIPT channel model and the coordinate system.}
\vspace{-15pt}
\end{figure}
Assuming non-dispersive narrow-band transmission and one-time reflection, the RIS-assisted OAM-SWIPT channel model is shown in Fig.~\ref{fig:channel_model}. The incident channel from the transmit UCA to the RIS, the reflected channel from the RIS to the receive UCA, and the LOS channel from the transmit UCA to the receive UCA are denoted by $\boldsymbol{\mathrm H}^{\rm In}\in \mathbb{C}^{N_I\times N_t}$, $\boldsymbol{\mathrm H}^{\rm Ref}\in \mathbb{C}^{N_r\times N_I}$, and $\boldsymbol{\mathrm H}^{\rm LOS}\in \mathbb{C}^{N_r\times N_t}$, respectively, where $N_I$ is the number of reflective elements in the RIS. In this paper, we use the uniform planar array (UPA) as the RIS. Each row of the RIS has $N^r_I$ elements, each column has $N^c_I$ elements, and $N_I\hspace{-0.1cm}=\hspace{-0.1cm}N^r_IN^c_I$. $\boldsymbol{\mathrm H}^{\rm In}$, $\boldsymbol{\mathrm H}^{\rm Ref}$, and $\boldsymbol{\mathrm H}^{\rm LOS}$ can be calculated according to the geometrical distribution of UCAs and RIS elements.

To make it easier to illustrate, we first establish the coordinate system as in Fig.~\ref{fig:channel_model2}. The $x$-axis is set from the center of the transmit UCA to the center of the first transmit antenna elements; the $z$-axis is set along the axis of the transmit UCA towards the transmission direction; the $y$-axis is given following the right-hand spiral rule. The center of the transmit UCA is located at the coordinate origin while the center of the receive UCA is located at $(d_x,d_y,D)$. $\theta_x$ and $\theta_y$ denote the angle between the transmit and the receive UCAs' normal lines along the $x$ and $y$ axes, respectively, with $0\hspace{-0.1cm}\le\hspace{-0.1cm}\theta_x, \theta_y\hspace{-0.1cm}\le\hspace{-0.1cm}\pi/2$. The deflection angle between the transmit and the receive UCAs' normal lines, denoted by $\theta$, can be given by $\theta\hspace{-0.1cm}=\hspace{-0.1cm}\arctan\sqrt{\tan^2\theta_x+\tan^2\theta_y}$. %We number the transmit UCA elements, of which the center is in the positive $x$-axis, the $1$st transmit element. The receive element, which is opposite to the $1$st transmit element, is set as the $1$st receive element. The rest transmit and receive elements are numbered from $2$ to $N_t$ and $N_r$ in a clockwise order viewed along $z$-axis, respectively.
%Due to severe channel attenuation, the RIS is deployed relatively close to the transmit UCA.
 The center of the RIS is located at $(p_x,p_y,p_z)$. $\theta^{R}_x$ and $\theta^{R}_y$ denote the angles between the transmit UCA and the RIS normal lines along the $x$ and $y$ axes, respectively, with $0\hspace{-0.1cm}\le\hspace{-0.1cm}\theta^{R}_x, \theta^{R}_y\hspace{-0.1cm}\le\hspace{-0.1cm}\pi/2$. $\theta^{R}\hspace{-0.1cm}=\hspace{-0.1cm}\arctan\sqrt{\tan^2\theta^{R}_x+\tan^2\theta^{R}_y}$ denotes the angle between the transmit UCA and the RIS normal lines.

Hence, the coordinates of the $n_t$th transmit element, the $n_r$th receive element, and the $n_I$th RIS element, denoted by $\boldsymbol{\mathrm p}^{t}_{n_t}$, $\boldsymbol{\mathrm p}^{r}_{n_r}$, and $\boldsymbol{\mathrm p}^{I}_{n_I}$, can be given as Eq.~\eqref{eq:coordinate}.
%\begin{subequations}
%\begin{numcases}{}
%\hspace{-0.1cm}\boldsymbol{\mathrm p}^{t}_{n_t} \hspace{-0.1cm}=\hspace{-0.1cm} \left[R_t\cos\left(\frac{2\pi \left(n_t-1\right)}{N_t}\right),R_t\sin\left(\frac{2\pi \left(n_t-1\right)}{N_t}\right),0\right];\\
%\label{eq:Mt}
%\end{numcases}
%\end{subequations}
In Eq.~\eqref{eq:coordinate}, ``${\rm mod}$'' is the modulo operation, ``$\Vert\cdot\Vert$'' denotes the two-norm operation, $\left(n_I\ |\ N^c_I\right)$ denotes $n_I$ divided by $N^c_I$, $d$ is the element spacing of the RIS, and
\begin{subequations}
\begin{numcases}{}
\boldsymbol{\mathrm a}^r = \left(\tan\theta_x,\tan\theta_y,1\right)\times(1,0,0);\\
\boldsymbol{\mathrm b}^r = \left(\tan\theta_x,\tan\theta_y,1\right)\times\boldsymbol{\mathrm a}^r;\\
\boldsymbol{\mathrm a}^I = \left(\tan\theta^{R}_x,\tan\theta^{R}_y,1\right)\times(1,0,0);\\
\boldsymbol{\mathrm b}^I = \left(\tan\theta^{R}_x,\tan\theta^{R}_y,1\right)\times\boldsymbol{\mathrm a}^I
\end{numcases}
\end{subequations}
are vectors perpendicular to the normal lines of the receive UCA and RIS. Here ``$\times$'' denotes the cross product. %Then, the coordinates of the transmit elements, the receive elements, and the RIS elements, denoted by $\boldsymbol{\mathrm P}^{t}\in \mathbb{C}^{N_t\times 3}$, $\boldsymbol{\mathrm P}^{r}\in \mathbb{C}^{N_r\times 3}$, and $\boldsymbol{\mathrm P}^{I}\in \mathbb{C}^{N_I\times 3}$, respectively, can be given as:
%\begin{subequations}
%\hspace{-0.1cm}\begin{numcases}{}
%\hspace{-0.1cm}\boldsymbol{\mathrm P}^{t} \hspace{-0.1cm}=\hspace{-0.1cm} \left[\left(\boldsymbol{\mathrm p}^{t}_{1}\right)^T,\cdots,\left(\boldsymbol{\mathrm p}^{t}_{n_t}\right)^T,\cdots,\left(\boldsymbol{\mathrm p}^{t}_{N_t}\right)^T\right]^T;\\
%\hspace{-0.1cm}\boldsymbol{\mathrm P}^{r} \hspace{-0.1cm}=\hspace{-0.1cm} \left[\left(\boldsymbol{\mathrm p}^{r}_{1}\right)^T,\cdots,\left(\boldsymbol{\mathrm p}^{r}_{n_r}\right)^T,\cdots,\left(\boldsymbol{\mathrm p}^{r}_{N_r}\right)^T\right]^T;\\
%\hspace{-0.1cm}\boldsymbol{\mathrm P}^{I} \hspace{-0.1cm}=\hspace{-0.1cm} \left[\left(\boldsymbol{\mathrm p}^{I}_{1}\right)^T,\cdots,\left(\boldsymbol{\mathrm p}^{I}_{n_I}\right)^T,\cdots,\left(\boldsymbol{\mathrm p}^{I}_{N_I}\right)^T\right]^T.
%\end{numcases}
%\end{subequations}
Thus, the $n_I$th row and the $n_t$th column of $\boldsymbol{\mathrm H}^{\rm In}$ (denoted by $H^{\rm In}_{n_I,n_t}$), the $n_r$th row and the $n_I$th column of $\boldsymbol{\mathrm H}^{\rm Ref}$ (denoted by $H^{\rm Ref}_{n_r,n_I}$), as well as the $n_r$th row and the $n_t$th column of $\boldsymbol{\mathrm H}^{\rm LOS}$ (denoted by $H^{\rm LOS}_{n_r,n_t}$), can be given as follows:
\begin{subequations}
\begin{numcases}{}
H^{\rm In}_{n_I,n_t} = \beta\frac{\lambda}{4\pi \Vert\boldsymbol{\mathrm p}^{I}_{n_I}-\boldsymbol{\mathrm p}^{t}_{n_t}\Vert} e^{-j\frac{2\pi}{\lambda}\Vert\boldsymbol{\mathrm p}^{I}_{n_I}-\boldsymbol{\mathrm p}^{t}_{n_t}\Vert};\\
H^{\rm Ref}_{n_r,n_I} = \beta\frac{\lambda}{4\pi \Vert\boldsymbol{\mathrm p}^{r}_{n_r}-\boldsymbol{\mathrm p}^{I}_{n_I}\Vert} e^{-j\frac{2\pi}{\lambda}\Vert\boldsymbol{\mathrm p}^{r}_{n_r}-\boldsymbol{\mathrm p}^{I}_{n_I}\Vert};\\
H^{\rm LOS}_{n_r,n_t} = \beta\frac{\lambda}{4\pi \Vert\boldsymbol{\mathrm p}^{r}_{n_r}-\boldsymbol{\mathrm p}^{t}_{n_t}\Vert} e^{-j\frac{2\pi}{\lambda}\Vert\boldsymbol{\mathrm p}^{r}_{n_r}-\boldsymbol{\mathrm p}^{t}_{n_t}\Vert},
\end{numcases}
\end{subequations}
where $\beta$ is a constant relative to the antenna element patterns and $\lambda$ represents the carrier wavelength. Therefore, the RIS-assisted OAM-SWIPT channel, denoted by $\boldsymbol{\mathrm H}\in \mathbb{C}^{N_r\times N_t}$, can be given as follows:
\begin{align}
\boldsymbol{\mathrm H} =
\hspace{-0.1cm}\sqrt{K}\boldsymbol{\mathrm H}^{\rm LOS}+\boldsymbol{\mathrm H}^{\rm Ref}\boldsymbol{\mathrm \Phi}\boldsymbol{\mathrm H}^{\rm In},
\label{eq:H}
\end{align}
where $K\in[0,1]$ denotes the attenuation factor of the LOS path, $\boldsymbol{\mathrm \Phi}\hspace{-0.1cm}=\hspace{-0.1cm}{\rm diag}\left(\boldsymbol{\mathrm \phi}\right)$ denotes the reflection-coefficient matrix of the RIS, ``${\rm diag}(\cdot)$'' denotes creating a diagonal matrix by using the elements of a vector, and $\boldsymbol{\mathrm \phi}\hspace{-0.1cm}=\hspace{-0.1cm} \left[e^{j\varphi_1},e^{j\varphi_2},\cdots,e^{j\varphi_{N_I}}\right]^T$ denotes the reflection coefficients corresponding to the RIS elements. We also denote by $\phi_{n_I}$ the $n_I$th element of $\boldsymbol{\mathrm \phi}$ with $|\phi_{n_I}| = 1$ and $1\hspace{-0.1cm}\le\hspace{-0.1cm}n_I\hspace{-0.1cm}\le\hspace{-0.1cm}N_I$. $\boldsymbol{\mathrm \phi}$ can be adjusted by the controller.

\section{RIS-assisted OAM Transmission Scheme}\label{sec:Transmission_scheme}
The multi-stream RF information signal is denoted by $\boldsymbol{\mathrm x}\in \mathbb{C}^{N_t\times 1}$ with $\mathbb{E}\left\{\boldsymbol{\mathrm x}\boldsymbol{\mathrm x}^H\right\}\hspace{-0.1cm}=\hspace{-0.1cm}\boldsymbol{\mathrm I}$, where $\boldsymbol{\mathrm I}$ is the identity matrix. $\boldsymbol{\mathrm x}$ is first modulated into the antenna excitations, denoted by $\boldsymbol{\mathrm s}\in \mathbb{C}^{N_t\times 1}$. The antenna excitations are with equal amplitudes and linearly increasing phases of $2\pi l/N_t$ for different OAM-modes, where $l$ represents the index of OAM-mode with $0\le l\le N_t-1$. This modulation is equivalent to doing unit inverse discrete Fourier transform (IDFT) for $\boldsymbol{\mathrm x}$, given as follows:
\begin{align}
\boldsymbol{\mathrm s} = \boldsymbol{\mathrm W}\boldsymbol{\mathrm P}_t\boldsymbol{\mathrm x},
\end{align}
where $\boldsymbol{\mathrm P}_t\hspace{-0.1cm}=\hspace{-0.1cm}{\rm diag}\left\{\left[\sqrt{P_0}, \sqrt{P_1}, \cdots, \sqrt{P_l}, \cdots,\sqrt{P_{N_t-1}}\right]\right\}$ is the power-allocating matrix, $P_l$ denotes the power allocated to the $l$th OAM-mode, and $\boldsymbol{\mathrm W}\in \mathbb{C}^{N_t\times N_t}$ is the unit IDFT matrix with the $n_1$th row and the $n_2$th column element given by ${W}_{n_1,n_2}\hspace{-0.1cm}=\hspace{-0.1cm}\frac{1}{\sqrt N_t}{\rm exp}[j{2\pi (n_1-1)(n_2-1)}/{N_t}]$.

%$\boldsymbol{\mathrm P}_t$ is a diagonal matrix, $P_t\hspace{-0.1cm}=\hspace{-0.1cm}\sum_{l=0}^{N_t-1}P_l$ denotes the total transmit power per symbol, and $\boldsymbol{\mathrm W}\in \mathbb{C}^{N_t\times N_t}$ is the unit IDFT matrix with the $n_1$th row and the $n_2$th column element given by ${W}_{n_1,n_2}\hspace{-0.1cm}=\hspace{-0.1cm}\frac{1}{\sqrt N_t}{\rm exp}[j{2\pi (n_1-1)(n_2-1)}/{N_t}]$. %We denote by $\boldsymbol{\mathrm w}_{n_2}$ the $n_2$th column of $\boldsymbol{\mathrm W}$.

%\begin{figure}[htbp]
%\centering
%%\vspace{-10pt}
%\includegraphics[scale=0.8]{channel_model.eps}
%%\vspace{-15pt}
%\caption{RIS-assisted OAM-SWIPT channel model.} \label{fig:channel_model}
%%\vspace{-5pt}
%\end{figure}

After passing through the RIS-assisted OAM-SWIPT channel, the received OAM signals, denoted by $\boldsymbol{\mathrm r}\in \mathbb{C}^{N_r\times 1}$, is given as follows:
\begin{align}
\boldsymbol{\mathrm r} &= \boldsymbol{\mathrm H}\boldsymbol{\mathrm s} + \boldsymbol{\mathrm n}
= \boldsymbol{\mathrm H}\boldsymbol{\mathrm W}\boldsymbol{\mathrm P}_t\boldsymbol{\mathrm x} + \boldsymbol{\mathrm n},
\end{align}
where $\boldsymbol{\mathrm n}\in \mathbb{C}^{N_r\times 1}$ denotes the narrow-band additive white Gaussian noise (AWGN) with independent and identically distributed elements following $\mathcal{CN}(0,\sigma_n^2)$. Next, the received signals are demodulated into the signals corresponding to different OAM-modes, denoted by $\boldsymbol{\mathrm y}$, using unit discrete Fourier transform (DFT) given as follows:
\begin{align}
\boldsymbol{\mathrm y} &= \boldsymbol{\mathrm W}'\boldsymbol{\mathrm r}
= \boldsymbol{\mathrm W}'\boldsymbol{\mathrm H}\boldsymbol{\mathrm W}\boldsymbol{\mathrm P}_t\boldsymbol{\mathrm x} + \boldsymbol{\mathrm W}'\boldsymbol{\mathrm n},
 \label{eq:y0}
\end{align}
where $\boldsymbol{\mathrm W}'\in \mathbb{C}^{N_r\times N_r}$ is the unit DFT matrix with the $n_1$th row and the $n_2$th column element given by ${W}'_{n_1,n_2}\hspace{-0.1cm}=\hspace{-0.1cm}\frac{1}{\sqrt N_r}{\rm exp}[-j{2\pi (n_1-1)(n_2-1)}/{N_r}]$. %We denote by $\boldsymbol{\mathrm w}'_{n_2}$ the $n_1$th row of $\boldsymbol{\mathrm W}'$.
%Since $\boldsymbol{\mathrm W}'$ is an unitary matrix, Eq.~\eqref{eq:y0} can be simplified as follows:
%\begin{align}
%\boldsymbol{\mathrm y} &= \boldsymbol{\mathrm W}'\boldsymbol{\mathrm H}\boldsymbol{\mathrm W}\boldsymbol{\mathrm P}_t\boldsymbol{\mathrm x} + \boldsymbol{\mathrm n},
% \label{eq:y}
%\end{align}

Finally, the signal of each OAM-mode splits into the ID stream and the EH stream. The power ratio of the ID stream to the EH stream is set as $\rho_l$ to $1-\rho_l$, where $0\hspace{-0.1cm}\le\hspace{-0.1cm} \rho_l \hspace{-0.1cm}\le\hspace{-0.1cm} 1$ and $t$ denotes the time. The power splitting ratio for different OAM-modes can also be denoted by a vector as $\boldsymbol{\rho}=[\rho_0,\rho_1,\cdots,\rho_{N_t-1}]$. For simplifying the following description, let $\boldsymbol{\mathrm H}^{\rm {OAM}}=\boldsymbol{\mathrm W}'\boldsymbol{\mathrm H}\boldsymbol{\mathrm W}$ denotes the OAM channel matrix with its $l_1$th row and the $l_2$th column element given by $H^{\rm OAM}_{l_1,l_2}$ denoting the complex channel from the $l_2$ transmitted OAM-mode to the $l_1$ received OAM-mode. Therefore, the recovered input signal, denoted by $\tilde{\boldsymbol{\mathrm x}}$, can be given as follows:
\begin{align}
&\tilde{\boldsymbol{\mathrm x}} \hspace{-0.1cm}=\hspace{-0.1cm} \mathop{\arg\min}\limits_{\boldsymbol{\mathrm x}\in \mathbb{C}_{N_t}}\hspace{-0.1cm}\left\Arrowvert{{\rm diag}\left(\boldsymbol{\mathrm h}^{\rm {OAM}}\right)^{\dag}\hspace{-0.1cm}\left[{\rm diag}\left(1\hspace{-0.1cm}-\hspace{-0.1cm}\boldsymbol{\rho}\right)\boldsymbol{\mathrm y} \hspace{-0.1cm}+\hspace{-0.1cm} \boldsymbol{\mathrm n}_{\rm cov}\right] \hspace{-0.1cm}-\hspace{-0.1cm} \boldsymbol{\mathrm P}_t\boldsymbol{\mathrm x}}\right\Arrowvert^2,
\label{eq:x_re}
\end{align}
where $\boldsymbol{\mathrm h}^{\rm {OAM}}$ denotes the vector formed by the diagonal elements of $\boldsymbol{\mathrm H}^{\rm {OAM}}$ and $\boldsymbol{\mathrm n}_{\rm cov}\in \mathbb{C}^{N_r\times 1}$ denotes the RF-band-to-baseband conversion AWGN with independent and identically distributed elements following $\mathcal{CN}(0,\sigma_{\rm cov}^2)$. By choosing specific values of $\boldsymbol{\rho}$, set suitable transmit power, and adjust the reflection-coefficient matrix of the RIS correspondingly, the maximum data rate can be achieved with minimum EH requirement.

%\addtolength{\textheight}{-0.44 cm}
\section{Problem Formulation}\label{sec:Problem_formulation}
Based on Eq.~\eqref{eq:x_re}, the signal-to-interference-plus-noise ratio (SINR) of the $l$th OAM-mode signal, denoted by $\gamma_l$ can be given as follows:
\begin{align}
\gamma_l = \frac{\left(1-\rho_l\right)\left|H^{\rm OAM}_{l,l}\right|^2P_l}{\left(1-\rho_l\right)\left(\sigma_n^2+\sum_{l'=0,l'\ne l}^{N_t-1}\left|H^{\rm OAM}_{l,l'}\right|^2P_{l'}\right)+\sigma_{\rm cov}^2}.
\label{eq:SINR}
\end{align}
Hence, the sum achievable rate of all OAM-modes, denoted by $C$, can be given as $\small{C \hspace{-0.1cm}=\hspace{-0.1cm} \sum_{l=0}^{N_t-1}{\rm log}\left(1+\gamma_l\right)}$.
%\begin{align}
%C &= \sum_{l=0}^{N_t-1}{\rm log}\left(1+\gamma_l\right)
%%\nonumber\\&= \sum_{l=0}^{N_t-1}{\rm log}\left(1+\frac{\left(1-\rho_l\right)\left|H^{\rm OAM}_{l,l}\right|^2P_l}{\left(1-\rho_l\right)\left(\sigma_n^2+\sum_{l'=0,l'\ne l}^{N_t-1}\left|H^{\rm OAM}_{l,l'}\right|^2P'_l\right)+\sigma_{\rm cov}^2}\right)
%.
%\label{eq:C}
%\end{align}
%Also, based on Eq.~\eqref{eq:SINR}, for OAM signals using $M$-ary quadrature amplitude modulation (QAM), the symbol error rate (SER), denoted by $P_e$, can be given as follows:
%\begin{align}
%P_e = \frac{4\left(\sqrt{M}-1\right)}{N_t\sqrt{M}}\sum_{l=0}^{N_t-1}{\rm erfc}\left(\sqrt{\frac{3\gamma_l}{M-1}}\right),
%\label{eq:Pe}
%\end{align}
%where erfc$(\alpha)=\frac{2}{\sqrt{\pi}}$$\int_{\alpha}^{\infty} e^{-t^2}dt$.
Then, the sum EH power, denoted by $Q$, can be given as $Q \hspace{-0.1cm}=\hspace{-0.1cm} \eta\sum_{l=0}^{N_t-1}\rho_l\left(\sigma_n^2+\sum_{l'=0}^{N_t-1}\left|H^{\rm OAM}_{l,l'}\right|^2P_{l'}\right)$,
%\begin{align}
%Q &= \eta\sum_{l=0}^{N_t-1}\rho_l\left(\sigma_n^2+\sum_{l'=0}^{N_t-1}\left|H^{\rm OAM}_{l,l'}\right|^2P_{l'}\right),
%\label{eq:Q}
%\end{align}
where $\eta$ denotes the harvested energy conversion efficiency and $0\le \eta\le 1$. %Let $\boldsymbol{ \gamma}\hspace{-0.1cm}=\hspace{-0.1cm}\left[\gamma_0,\gamma_1,\cdots,\gamma_{N_t-1}\right]^{T}$ denotes the SINR vector, Eqs.~\eqref{eq:C} to \eqref{eq:Q} can be rewritten in matrix form as follows:

Therefore, for jointly optimizing the transmit power $\boldsymbol{\mathrm P}_t$, the reflection
coefficients $\boldsymbol{\mathrm \phi}$, and the power split ratio $\boldsymbol{\rho}$ to maximize the sum achievable capacity $C$ subject to the EH requirement and the given total power budget, the problem can be formulated as follows:
\begin{subequations}\label{eq:P1}
\begin{align}
\boldsymbol{\mathrm P1:}\max_{\boldsymbol{\mathrm P}_t,\boldsymbol{\mathrm \phi},\boldsymbol{\rho}} \quad &C\\
{\mathrm {s.t.}} \quad
&Q\ge \bar{Q}, \\
&0\le P_t \le \bar{P_t}, \\
&\rho_l\in [0,1],|\phi_{n_I}| = 1,
\end{align}
\end{subequations}
where $P_t\hspace{-0.1cm}=\hspace{-0.1cm}\sum_{l=0}^{N_t-1}P_l$ denotes the total transmit power, $\bar{P_t}$ denotes the maximum transmit power budget, and $\bar{Q}$ denotes the minimum EH requirement.

Since the purpose of this paper is to provide a kind of benchmark that verifies the feasibility and capacity enhancement of combining RIS with OAM-SWIPT, we simplify $\boldsymbol{\mathrm P1}$ by evenly allocating the transmit power to each OAM-mode and setting $P_t\hspace{-0.1cm}=\hspace{-0.1cm}\bar{P_t}$ to maximize the achievable capacity. Thus, $\boldsymbol{\mathrm P1}$ can be simplified as follows:
\begin{subequations}\label{eq:P2}
\begin{align}
\boldsymbol{\mathrm P2:}\max_{\boldsymbol{\mathrm \phi},\boldsymbol{\rho}} \quad &\sum_{l=0}^{N_t-1}{\rm log}\left(1+\gamma'_l\right)\\
{\mathrm {s.t.}} \quad
&\eta\sum_{l=0}^{N_t-1}\rho_l\left(\sigma_n^2+\frac{\bar{P}_t}{N_t}\sum_{l'=0}^{N_t-1}\left|H^{\rm OAM}_{l,l'}\right|^2\right)\ge\bar{Q}, \\
&\rho_l \in [0,1],|\phi_{n_I}| = 1,
\end{align}
\end{subequations}
where
\begin{align}
\gamma'_l = \frac{\frac{\bar{P_t}}{N_t}\left(1-\rho_l\right)\left|H^{\rm OAM}_{l,l}\right|^2}{\left(1-\rho_l\right)\left(\sigma_n^2+\frac{\bar{P_t}}{N_t}\sum_{l'=0,l'\ne l}^{N_t-1}\left|H^{\rm OAM}_{l,l'}\right|^2\right)+\sigma_{\rm cov}^2}.
\end{align}

\section{Alternating Optimization Algorithm}\label{sec:Algorithm}
To solve $\boldsymbol{\mathrm P2}$, we divide it into two subproblems which optimize $\boldsymbol{\mathrm \phi}$ with fixed $\boldsymbol{\mathrm \rho}$ and optimize $\boldsymbol{\mathrm \rho}$ with fixed $\boldsymbol{\mathrm \phi}$, respectively. We then alternatively solve them until convergence. We assume that the receiver knows perfect channel state information.

\subsection{Optimizing the Reflection Coefficients $\boldsymbol{\mathrm \phi}$}
Fixing $\boldsymbol{\mathrm \rho}$ and ignoring the EH requirement, $\boldsymbol{\mathrm P2}$ can be reformulated as follows:
\begin{subequations}\label{eq:SP1}
\begin{align}
\boldsymbol{\mathrm P3:}\max_{\boldsymbol{\mathrm \phi}} \quad &\sum_{l=0}^{N_t-1}{\rm log}\left(1+\gamma'_l\right)\\
{\mathrm {s.t.}} \quad
&|\phi_{n_I}| = 1.
\end{align}
\end{subequations}

By introducing a positive semidefinite auxiliary matrix $\boldsymbol{\mathrm F}\in \mathbb{C}^{N_t\times N_t}$, we use the weighted minimum mean square error (WMMSE) method to reformulate $\boldsymbol{\mathrm P3}$ as follows\cite{IRS_MIMO,oam_irs}:
\begin{subequations}\label{eq:SP1_1}
\begin{align}
\boldsymbol{\mathrm P4:}\max_{\boldsymbol{\mathrm \phi},\boldsymbol{\mathrm F}} \quad &{\rm log}\left|\boldsymbol{\mathrm F}\right| - {\mathrm tr}\left(\boldsymbol{\mathrm F}\boldsymbol{\mathrm E}\right) + N_t\\
{\mathrm {s.t.}} \quad
&|\phi_{n_I}| = 1.
\end{align}
\end{subequations}
where ``${\mathrm tr}\left(\cdot\right)$'' is the trace operation, and $\boldsymbol{\mathrm E}\in \mathbb{C}^{N_t\times N_t}$ denotes the mean square error (MSE) matrix of the OAM transmission. Based on Eq.~\eqref{eq:x_re}, $\boldsymbol{\mathrm E}$ can be given as follows:
\begin{align}
\boldsymbol{\mathrm E} \hspace{-0.1cm}=\hspace{-0.1cm}& \left[{\rm diag}(\boldsymbol{\mathrm h}^{\rm {OAM}})^{\dag}\boldsymbol{\mathrm H}^{\rm {OAM}} - \boldsymbol{\mathrm I}\right]\left[{\rm diag}(\boldsymbol{\mathrm h}^{\rm {OAM}})^{\dag}\boldsymbol{\mathrm H}^{\rm {OAM}} - \boldsymbol{\mathrm I}\right]^{H}\nonumber\\
&+ {\rm diag}(\boldsymbol{\mathrm h}^{\rm {OAM}})^{\dag}\left[{\rm diag}(\boldsymbol{\mathrm h}^{\rm {OAM}})^{\dag}\right]^H\frac{N_t}{\bar{P}_t}\left(\sigma_n^2+\frac{\sigma_{\rm cov}^2}{\left(1-\boldsymbol{\mathrm \rho}\right)}\right),
\end{align}
where ``$(\cdot)^H$'' is the conjugate transpose operation. Since $\boldsymbol{\mathrm F}$ only appears in the objective function, the optimal solution of $\boldsymbol{\mathrm F}$ can be obtained by setting the first-order derivative of the objective function with respect to $\boldsymbol{\mathrm F}$ to zero. Thus, the optimal solution of $\boldsymbol{\mathrm F}$ is denoted by $\boldsymbol{\mathrm F}^* \hspace{-0.1cm}=\hspace{-0.1cm} \left(\boldsymbol{\mathrm E}^{\rm MMSE}\right)^{-1}$, where $\boldsymbol{\mathrm E}^{\rm MMSE}$ denotes the MSE matrix of the MMSE receiver.
%given as follows:
%\begin{align}
%\boldsymbol{\mathrm E}^{\rm MMSE} = \boldsymbol{\mathrm I}_{N_t} - \boldsymbol{\mathrm G}^H\left(\sigma^2_n\left(\boldsymbol{\mathrm I}_{N_r}-{\rm diag}\left(\boldsymbol{\mathrm \rho}\right)\right) + \sigma^2_{\rm cov}\boldsymbol{\mathrm I}_{N_r}\right)^{-1}\boldsymbol{\mathrm G}.
% \label{eq:MSE_MMSE}
%\end{align}

After removing the terms that are independent of $\boldsymbol{\mathrm \phi}$, $\boldsymbol{\mathrm P4}$ can be rewritten as follows:
\begin{subequations}\label{eq:SP1_2}
\begin{align}
\boldsymbol{\mathrm P5:}\min_{\boldsymbol{\mathrm \phi}} \quad &{\mathrm tr}\left[\boldsymbol{\mathrm F}\boldsymbol{\mathrm H}^{\rm {OAM}}\left(\boldsymbol{\mathrm H}^{\rm {OAM}}\right)^H\right]\nonumber\\ &- {\mathrm tr}\left(\boldsymbol{\mathrm F}\boldsymbol{\mathrm H}^{\rm {OAM}}\right) - {\mathrm tr}\left[\boldsymbol{\mathrm F}\left(\boldsymbol{\mathrm H}^{\rm {OAM}}\right)^H\right]\\
{\mathrm {s.t.}} \quad
&|\phi_{n_I}| = 1.
\end{align}
\end{subequations}
By substituting Eq.~\eqref{eq:H} into $\boldsymbol{\mathrm P5}$ and defining
\begin{subequations}
\begin{numcases}{}
\boldsymbol{\mathrm B} = \left(\boldsymbol{\mathrm H}^{\rm Ref}\right)^H\left(\boldsymbol{\mathrm W}'\right)^H\boldsymbol{\mathrm F}\boldsymbol{\mathrm W}'\boldsymbol{\mathrm H}^{\rm LOS}\left(\boldsymbol{\mathrm H}^{\rm LOS}\right)^H;\\
\boldsymbol{\mathrm C} = \boldsymbol{\mathrm H}^{\rm In}\left(\boldsymbol{\mathrm H}^{\rm LOS}\right)^H\left(\boldsymbol{\mathrm W}'\right)^H\boldsymbol{\mathrm F}\boldsymbol{\mathrm W}'\boldsymbol{\mathrm H}^{\rm Ref};\\
\boldsymbol{\mathrm D} = \boldsymbol{\mathrm H}^{\rm In}\left(\boldsymbol{\mathrm H}^{\rm In}\right)^H;\\
\boldsymbol{\mathrm J} = \boldsymbol{\mathrm H}^{\rm In}\boldsymbol{\mathrm F}\boldsymbol{\mathrm W'}\boldsymbol{\mathrm H}^{\rm Ref},
\end{numcases}
\end{subequations}
$\boldsymbol{\mathrm P5}$ can be further simplified as follows:
%\begin{subequations}\label{eq:SP1_3}
%\begin{align}
%\min_{\boldsymbol{\mathrm \phi}} \quad\quad &{\mathrm tr}\left(\boldsymbol{\mathrm \Phi}^H\boldsymbol{\mathrm B}\boldsymbol{\mathrm \Phi}\boldsymbol{\mathrm C}\right)\nonumber\\ &+ {\mathrm tr}\left[\boldsymbol{\mathrm \Phi}^H\left(\boldsymbol{\mathrm D}-\boldsymbol{\mathrm G}\right)^H\right] + {\mathrm tr}\left[\boldsymbol{\mathrm \Phi}\left(\boldsymbol{\mathrm D}-\boldsymbol{\mathrm G}\right)\right]\\
%{\mathrm {s.t.}} \quad\quad
%&|\phi_{n_I}| = 1.
%\end{align}
%\end{subequations}
\begin{subequations}\label{eq:SP1_3}
\begin{align}
\boldsymbol{\mathrm P6:}\min_{\boldsymbol{\mathrm \phi}} \quad &\boldsymbol{\mathrm \phi}^H\boldsymbol{\mathrm R}\boldsymbol{\mathrm \phi} + {\rm Re}\left\{\boldsymbol{\mathrm v}^T\boldsymbol{\mathrm \phi}\right\}\\
{\mathrm {s.t.}} \quad
&|\phi_{n_I}| = 1,
\end{align}
\end{subequations}
where $\boldsymbol{\mathrm R} \hspace{-0.1cm}=\hspace{-0.1cm} \boldsymbol{\mathrm B}\odot \boldsymbol{\mathrm D}^T$, ``$\odot$'' denotes the Hadamard product, $\boldsymbol{\mathrm v}\hspace{-0.1cm}=\hspace{-0.1cm}\left[{\rm Diag}\left(\boldsymbol{\mathrm C}-\boldsymbol{\mathrm J}\right)\right]$, and ``${\rm Diag}(\cdot)$'' denotes creating a vector with the diagonal elements of a matrix. $\boldsymbol{\mathrm P6}$ is a quadratically constrained quadratic program (QCQP) problem and can be equivalently reformulated as follows:
\begin{subequations}\label{eq:SP1_4}
\begin{align}
\boldsymbol{\mathrm P7:}\min_{\hat{\boldsymbol{\mathrm \phi}}} \quad &\hat{\boldsymbol{\mathrm \phi}}^H\hat{\boldsymbol{\mathrm R}}\hat{\boldsymbol{\mathrm \phi}}\\
{\mathrm {s.t.}} \quad
&|\hat{\phi}_{n_I}| = 1,|\hat{\phi}_{N_I+1}| = 1,
\end{align}
\end{subequations}
where $\hat{\boldsymbol{\mathrm \phi}}\hspace{-0.1cm}=\hspace{-0.1cm}
\left[ {\begin{array}{*{20}{c}}
\boldsymbol{\mathrm \phi}\\
t
\end{array}}\right]$ and $\hat{\boldsymbol{\mathrm R}}\hspace{-0.1cm}=\hspace{-0.1cm}
\left[ {\begin{array}{*{20}{c}}
\boldsymbol{\mathrm R}&\frac{1}{2}\boldsymbol{\mathrm v}^*\\
\frac{1}{2}\boldsymbol{\mathrm v}^T&0
\end{array}}\right]$. By defining $\hat{\boldsymbol{\mathrm \Phi}}\hspace{-0.1cm}=\hspace{-0.1cm}\hat{\boldsymbol{\mathrm \phi}}\hat{\boldsymbol{\mathrm \phi}}^H$ and applying semidefinite relaxation (SDR), $\boldsymbol{\mathrm P7}$ can be transformed into a standard convex semidefinite problem as follows\cite{IRS_op}:
\begin{subequations}\label{eq:SP1_5}
\begin{align}
\boldsymbol{\mathrm P8:}\min_{\hat{\boldsymbol{\mathrm \Phi}}} \quad &{\rm tr}\left(\hat{\boldsymbol{\mathrm R}}\hat{\boldsymbol{\mathrm \Phi}}\right)\\
{\mathrm {s.t.}} \quad
&|\hat{\Phi}_{n_I,n_I}| = 1,|\hat{\Phi}_{N_I+1,N_I+1}| = 1,\hat{\boldsymbol{\mathrm \Phi}}\succeq0,
\end{align}
\end{subequations}
which can be optimally solved by existing convex optimization solvers\cite{cvx}. After getting the optimal $\hat{\boldsymbol{\mathrm \Phi}}$, denoted by $\hat{\boldsymbol{\mathrm \Phi}}^{\rm op}$, we obtain the eigenvalue decomposition of $\hat{\boldsymbol{\mathrm \Phi}}^{\rm op}$, given as $\hat{\boldsymbol{\mathrm \Phi}}^{\rm op}\hspace{-0.1cm}=\hspace{-0.1cm}\boldsymbol{\mathrm U}\boldsymbol{\mathrm \Sigma}\boldsymbol{\mathrm U}^H$, where $\boldsymbol{\mathrm U}\in \mathbb{C}^{(N_I+1)\times (N_I+1)}$ is a unitary matrix and $\boldsymbol{\mathrm \Sigma}\in \mathbb{C}^{(N_I+1)\times (N_I+1)}$ is a diagonal matrix. By introducing a random vector with its independent and identically distributed elements following $\mathcal{CN}(0,1)$, denoted by $\boldsymbol{\mathrm \zeta}\in \mathbb{C}^{(N_I+1)\times 1}$, the optimal solution to $\boldsymbol{\mathrm P7}$, denoted by $\hat{\boldsymbol{\mathrm \phi}}^{\rm op}$ can be given as the maximum one of $\hat{\boldsymbol{\mathrm \phi}}\hspace{-0.1cm}=\hspace{-0.1cm}\boldsymbol{\mathrm U}\boldsymbol{\mathrm \Sigma}^{1/2}\boldsymbol{\mathrm \zeta}$ with different $\boldsymbol{\mathrm \zeta}$'s. Thus, the optimal $\boldsymbol{\mathrm \phi}$, denoted by $\boldsymbol{\mathrm \phi}^{\rm op}$, can be given as follows:
\begin{align}
\boldsymbol{\mathrm \phi}^{\rm op} = e^{j{\rm arg}\left(\left[\frac{\hat{\boldsymbol{\mathrm \phi}}^{\rm op}}{\hat{\mathrm \phi}^{\rm op}_{N_I+1}}\right]_{(1:N_I)}\right)},
\label{eq:phiop}
\end{align}
where $\hat{\mathrm \phi}_{N_I+1}$ is the $N_I+1$th element of $\hat{\boldsymbol{\mathrm \phi}}$ and ``$[\cdot]_{(1:N_I)}$'' denotes the first $N_I$ elements of a vector.
%where
%\begin{align}
%\gamma''_l = \frac{\frac{\bar{P_t}}{N_t}\left(1-\rho_l\right)\left|\boldsymbol{\mathrm \phi}\bar{\boldsymbol{\mathrm w}}_{l,l}\right|^2}{{\left(1-\rho_l\right)\left(\sigma_n^2+\frac{\bar{P_t}}{N_t}\sum_{l'=0,l'\ne l}^{N_t-1}\left|\boldsymbol{\mathrm \phi}\boldsymbol{\bar{\mathrm w}}_{l,l'}\right|^2\right)+\sigma_{\rm cov}^2}}
%\end{align}
%and
%\begin{align}
%\bar{\boldsymbol{\mathrm w}}_{l_1,l_2} = {\rm diag}\left(\boldsymbol{\mathrm w'}_{l_1}\boldsymbol{\mathrm H}^{\rm Ref}\right)\boldsymbol{\mathrm H}^{\rm In}\boldsymbol{\mathrm w}_{l_2}.
%\end{align}

\subsection{Optimizing the Power Splitting Ratio $\boldsymbol{\mathrm \rho}$}
Fixing $\boldsymbol{\mathrm \phi}$, $\boldsymbol{\mathrm P2}$ can be reformulated as follows:
\begin{subequations}\label{eq:SP2}
\begin{align}
\boldsymbol{\mathrm P9:}\max_{\boldsymbol{\rho}} \quad &\sum_{l=0}^{N_t-1}{\rm log}\left(1+\gamma'_l\right)\\
{\mathrm {s.t.}} \quad
&\eta\sum_{l=0}^{N_t-1}\rho_l\left(\sigma_n^2+\frac{\bar{P}_t}{N_t}\sum_{l'=0}^{N_t-1}\left|H^{\rm OAM}_{l,l'}\right|^2\right)\ge\bar{Q}, \\
&\rho_l \in [0,1],
\end{align}
\end{subequations}
which is a standard convex semidefinite problem and can be optimally solved by existing convex optimization solvers.

%\subsection{Alternatively Solve $\boldsymbol{\mathrm P3}$ and $\boldsymbol{\mathrm P9}$}
The whole alternating algorithm for optimizing $\boldsymbol{\mathrm \phi}$ and $\boldsymbol{\mathrm \rho}$ in $\boldsymbol{\mathrm P2}$ is summarized in Algorithm~\ref{al:1}.
\begin{algorithm}
	%\textsl{}\setstretch{1.8}
	\renewcommand{\algorithmicrequire}{\textbf{Input:}}
	\renewcommand{\algorithmicensure}{\textbf{Output:}}
	\caption{: Alternatively optimize $\boldsymbol{\mathrm \phi}$ and $\boldsymbol{\mathrm \rho}$;}
	\label{al:1}
	\begin{algorithmic}[1]
		\STATE Initialization:$\boldsymbol{\mathrm \phi}^{(0)}=\boldsymbol{\mathrm 0}, \boldsymbol{\mathrm \rho}^{(0)}=\boldsymbol{\mathrm 0}, t=0$
		\REPEAT
		\STATE $t = t + 1$;
		\STATE Obtain $\hat{\boldsymbol{\mathrm \Phi}}^{\rm op}$ based on $\boldsymbol{\mathrm P8}$;
        \STATE Calculate the eigenvalue decomposition of $\hat{\boldsymbol{\mathrm \Phi}}^{\rm op}$ as $\hat{\boldsymbol{\mathrm \Phi}}^{\rm op}\hspace{-0.1cm}=\hspace{-0.1cm}\boldsymbol{\mathrm U}\boldsymbol{\mathrm \Sigma}\boldsymbol{\mathrm U}^H$;
        \FOR{$k=1:10000$}
        \STATE Randomly generate $\boldsymbol{\mathrm \zeta}$ and calculate $\hat{\boldsymbol{\mathrm \phi}}=\boldsymbol{\mathrm U}\boldsymbol{\mathrm \Sigma}_{\phi}^{1/2}\boldsymbol{\mathrm \zeta}$;
        \ENDFOR
        \STATE Choose one $\hat{\boldsymbol{\mathrm \phi}}$ that yields the best objective of $\boldsymbol{\mathrm P7}$ as $\hat{\boldsymbol{\mathrm \phi}}^{\rm op}$;
        \STATE Update $\boldsymbol{\mathrm \phi}^{(t)}$ based on Eq.~\eqref{eq:phiop};
        \IF {$Q< \bar{Q}$ for $\forall |\rho_l|\le1$}
        \STATE break;
        \ENDIF
		\STATE Update $\boldsymbol{\mathrm \rho}^{(t)}$ based on $\boldsymbol{\mathrm P9}$;
		\UNTIL convergence.
	\end{algorithmic}
\end{algorithm}

\section{Simulations}\label{sec:Simulation}
In this section, we first give simulations to evaluate the convergence of our developed algorithm. Then, the capacity and EH enhancements brought by the RIS for OAM-SWIPT are evaluated. Also, we compare OAM-SWIPT and MIMO-SWIPT. Besides, we simulate how the distance between the transmit UCA and the RIS impacts the capacity and EH performances of the RIS-assisted OAM-SWIPT system.
\subsection{Convergence Evaluation}\label{sec:Convergence}
\begin{figure}[htbp]
\centering
\vspace{-15pt}
\includegraphics[scale=0.5]{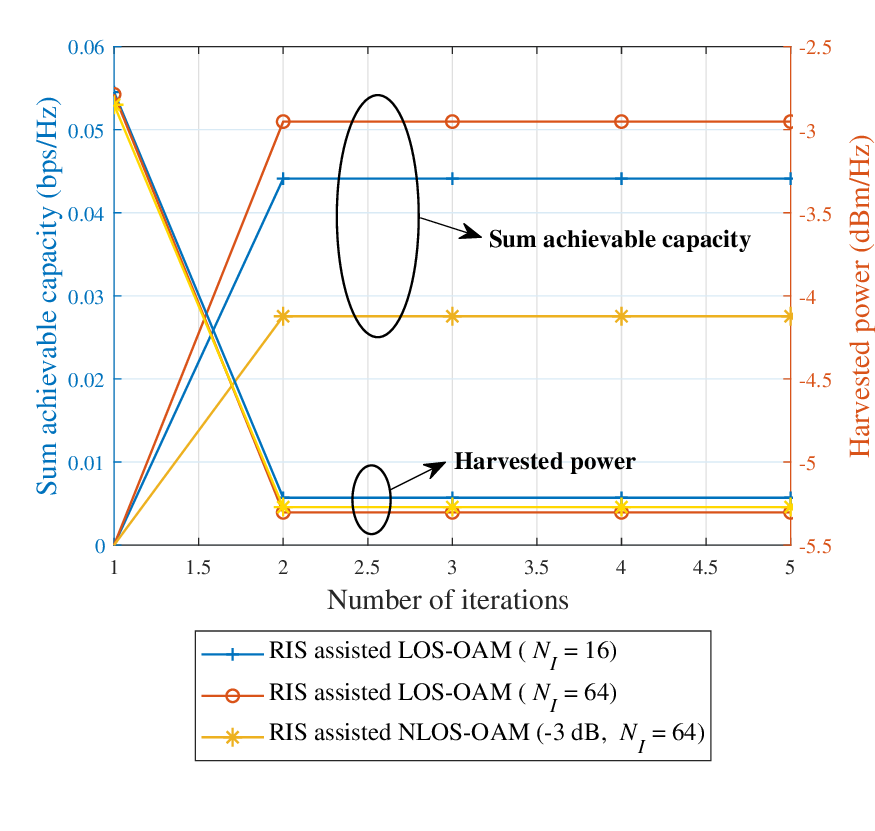}
\vspace{-20pt}
\caption{Iteration times of our developed algorithm.} \label{fig:Iteration_times_R}
\vspace{-10pt}
\end{figure}
We first evaluate the convergence of our proposed algorithm. In Fig.~\ref{fig:Iteration_times_R}, we set $\beta\hspace{-0.1cm}=\hspace{-0.1cm}1$, $\lambda\hspace{-0.1cm}=\hspace{-0.1cm}0.05$ m, $P_t\hspace{-0.1cm}=\hspace{-0.1cm}30$ dBm, $\sigma_n^2\hspace{-0.1cm}=\hspace{-0.1cm}-20$ dBm, $\sigma_{\rm cov}^2\hspace{-0.1cm}=\hspace{-0.1cm}-33$ dBm, $\eta\hspace{-0.1cm}=\hspace{-0.1cm}0.8$, $N_t\hspace{-0.1cm}=\hspace{-0.1cm}N_r\hspace{-0.1cm}=\hspace{-0.1cm}8$, $R_t\hspace{-0.1cm}=\hspace{-0.1cm}R_r\hspace{-0.1cm}=\hspace{-0.1cm}0.1$ m, $d_x\hspace{-0.1cm}=\hspace{-0.1cm}d_y\hspace{-0.1cm}=\hspace{-0.1cm}0$, $\theta_x\hspace{-0.1cm}=\hspace{-0.1cm}\theta_y\hspace{-0.1cm}=\hspace{-0.1cm}0$, $D\hspace{-0.1cm}=\hspace{-0.1cm}20$ m, $d\hspace{-0.1cm}=\hspace{-0.1cm}0.025$ m, $(p_x,p_y,p_z)\hspace{-0.1cm}=\hspace{-0.1cm}(0,-0.2,0.4)$ m, $\theta_x^R\hspace{-0.1cm}=\hspace{-0.1cm}0$, and $\theta_y^R\hspace{-0.1cm}=\hspace{-0.1cm}\pi/2$. Simulations are given by the $fmincon$ function in MATLAB based on our developed algorithm. We also set $K\hspace{-0.1cm}=\hspace{-0.1cm}1$ and $K\hspace{-0.1cm}=\hspace{-0.1cm}0.5$ as well as $N_I^r\hspace{-0.1cm}=\hspace{-0.1cm}N_I^c\hspace{-0.1cm}=\hspace{-0.1cm}4$ and $N_I^r\hspace{-0.1cm}=\hspace{-0.1cm}N_I^c\hspace{-0.1cm}=\hspace{-0.1cm}8$ to evaluate the convergence in different situations. Figure~\ref{fig:Iteration_times_R} shows that for all situations, at most three iterations are needed for convergence, which suggests a low implementation complexity of our algorithm.

\subsection{Capacity and EH Enhancements}\label{sec:Enhancement}
\begin{figure}[htbp]
\centering
\vspace{-10pt}
\includegraphics[scale=0.5]{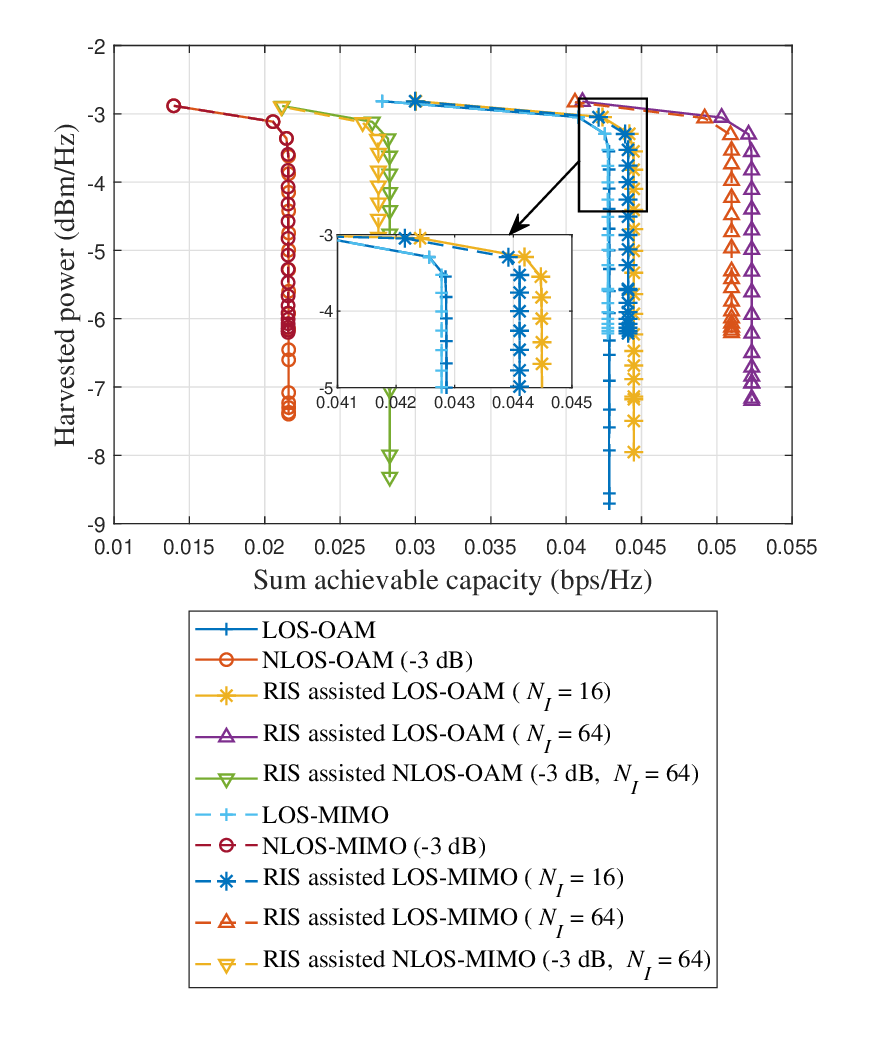}
\vspace{-20pt}
\caption{Capacity and EH enhancement brought by the RIS for the OAM-SWIPT system.} \label{fig:RIS_enhancement}
%\vspace{-10pt}
\end{figure}
Then, we evaluate the capacity and EH enhancements brought by the RIS for the OAM-SWIPT system. We also compare OAM-SWIPT and MIMO-SWIPT in this part. In Fig.~\ref{fig:RIS_enhancement}, we set the variables to remain the same as those in Fig.~\ref{fig:Iteration_times_R}. To evaluate the capacity and EH enhancements brought by the RIS, we plot the power-capacity performances of LOS-OAM and NLOS-OAM without the assistance of the RIS. To compare the capacities of OAM-SWIPT and MIMO-SWIPT, we also plot the power-capacity performances of MIMO-SWIPT in Fig.~\ref{fig:RIS_enhancement}. Figure~\ref{fig:RIS_enhancement} suggests that the RIS can increase both the capacity and EH power for the OAM-SWIPT systems. Furthermore, the OAM-SWIPT benefits from the RIS in both LOS and NLOS scenarios. Also, more RIS elements can bring higher capacity and EH enhancement. the MIMO-SWIPT systems can also benefit from the RIS. But the gap between MIMO and OAM-SWIPT gets bigger and bigger with the increase of RIS element number.

\subsection{Impact of the Distance Between the Transmit UCA and the RIS}\label{sec:RIS_Dis_UCA_Impact}
\begin{figure}[htbp]
\centering
%\vspace{-20pt}
\includegraphics[scale=0.5]{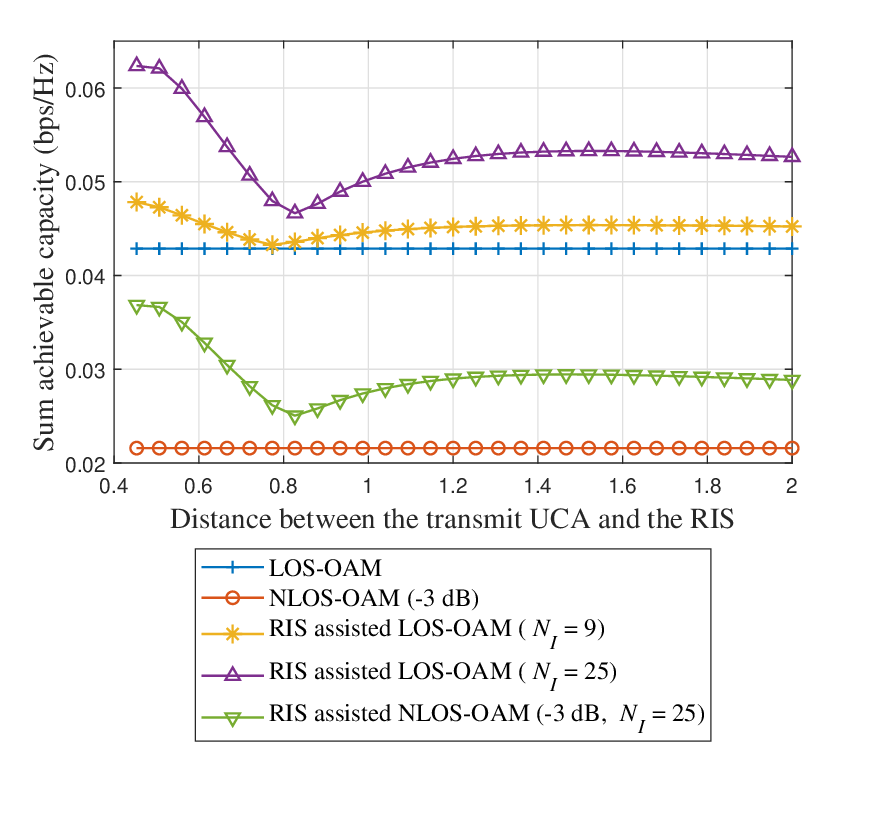}
\vspace{-20pt}
\caption{Impact of the distance between the transmit UCA and the RIS.} \label{fig:RIS_Dis_UCA_Impact}
\vspace{-10pt}
\end{figure}
In this subsection, we analyze how the distance between the transmit UCA and the RIS impacts the sum achievable capacity of the RIS-assisted OAM-SWIPT system. In Fig.~\ref{fig:RIS_Dis_UCA_Impact}, we change the distance between the transmit UCA and the RIS from $0.4$ to $2$ m. Other variables remain the same as those in Fig.~\ref{fig:RIS_enhancement}. Figure~\ref{fig:RIS_Dis_UCA_Impact} shows that the capacity first decreases sharply, then increases smoothly, and finally decreases more smoothly as the distance increases. Thus, the RIS should be deployed relatively close to the transmit antenna but avoid the performance bottom.

\section{Conclusions}\label{sec:Conclusion}
In this paper, we modeled and derived the optimal reflection coefficients and power splitting ratio for the RIS-assisted OAM-SWIPT transmission. We first proposed the system and channel models. Then, we proposed the RIS-assisted OAM transmission scheme. Based on the transmission model, we formulated the capacity maximization problem, which jointly optimized the reflection coefficients and the power splitting ratio of the ID stream to the EH stream to maximize the sum achievable rate under the minimum EH constraint. Next, we solved the problem by dividing it into two subproblems. Simulations validated the convergence of our developed algorithm. We also evaluated the capacity and EH enhancements brought by the RIS for the OAM-SWIPT system. Besides, we compared OAM-SWIPT and MIMO-SWIPT. Furthermore, the impact of the distance between the transmit UCA and the RIS is also simulated, which showed that the RIS should be deployed relatively close to the transmit antenna but avoid the performance bottom.

%\begin{appendices}
%\section{Proof for Theorem~\ref{the:channel_matrix}}\label{pro:multi-coil channel}
%
%\end{appendices}

\bibliographystyle{IEEEtran}
\bibliography{References}

\ifCLASSOPTIONcaptionsoff
  \newpage
\fi

\end{document}